%% file: neurips_2026.tex
\documentclass{article}

\PassOptionsToPackage{numbers, compress}{natbib}

\usepackage{float}
\usepackage[preprint]{neurips_2026}

\usepackage[utf8]{inputenc} 
\usepackage[T1]{fontenc}    
\usepackage{hyperref}       
\usepackage{url}            
\usepackage{booktabs}       
\usepackage{amsfonts}       
\usepackage{nicefrac}       
\usepackage{microtype}      
\usepackage{xcolor}         
\usepackage{amsmath}
\usepackage{graphicx} 
\usepackage{amssymb}
\usepackage{tabularx}
\usepackage{booktabs}
\usepackage{tabularx}
\usepackage{listings}
\usepackage{placeins}      
\usepackage{xcolor}
\usepackage{subcaption}

\usepackage[table]{xcolor}

\usepackage{booktabs}
\usepackage{makecell}

\definecolor{lightgrayrow}{gray}{0.92}
\definecolor{headergray}{gray}{0.85}
\usepackage{array}
\usepackage{amssymb}

\title{Jobs' AI Exposure Should Be Measured from Evidence, Not Model Priors}

\newcommand{\maybeincludegraphics}[2][]{%
  \IfFileExists{#2}{%
    \includegraphics[#1]{#2}%
  }{%
    \fbox{\parbox{0.9\linewidth}{\centering
      Missing figure asset: \texttt{\detokenize{#2}}%
    }}%
  }%
}
\author{%
  Luca Mouchel$^{1,2}$ \hspace{1cm} Pierre Bouquet$^{2}$ \hspace{1cm} Yossi Sheffi$^{2}$\\
  $^{1}$École Polytechnique Fédérale de Lausanne (EPFL)\\
  $^{2}$Massachusetts Institute of Technology (MIT)\\
  \texttt{\{lucamo02,pibou149,sheffi\}@mit.edu} \\
}

\begin{document}

\maketitle

\begin{abstract}
This position paper argues that job exposure to AI should be measured with grounded, evidence-based methods, not inferred from LLM priors alone. Current theoretical exposure measures use zero-shot prompting to classify task-level AI exposure, generating labels with no explicit evidence, no transparent chain of reasoning, and no external validation. 
The stakes of these measurements are too high to rely on such methods, as they influence policy making, where public and private funds are directed, and how workers understand their future prospects. We therefore argue that AI capability claims should meet three standards: reproducibility, external grounding, and inspectability.
We propose a retrieval-augmented framework that assigns AI exposure labels to all 18,796 occupation--task pairs in O*NET 30.2, using open-weight reasoning and instruct models with retrieved news articles and academic paper abstracts as evidence of current AI capabilities. Relative to a zero-shot baseline, the grounded condition is preferred in over 72\% of disagreement cases under both automatic and human evaluation, and yields scores that align more closely with observed real-world AI usage. Taken together, these findings suggest that evidence-grounded measurement better captures what current AI systems can plausibly do in practice, rather than what a model asserts without external evidence.
Because AI capabilities continue to change, the measurements used to inform policy must evolve with them: theoretical AI exposure scores should be periodically reassessed, not inherited as immutable ground truth.


\end{abstract}

\section{Introduction}
\input{1_intro_v3}

\section{Related Work}
\input{2_relatedwork}
\input{3_data}

\input{4_method}


\input{6_results}

\vspace{-10pt}
\section{Discussion, Limitations and Alternate Views}

\input{7_limitations}

\bibliographystyle{plainnat}
\bibliography{references}

\appendix

\input{8_appendix}


\newpage


\end{document}

%% file: 1_intro_v3.tex
\label{sec:intro}

Measures of occupational exposure to AI now shape economic analysis, workforce planning, and regulatory debate. International institutions, including the World Economic Forum, the International Monetary Fund, the International Labor Organization, and the OECD have each produced analyses of AI's labor-market impact that inform decisions about worker protection, reskilling, and preparedness for rapidly improving AI systems \citep{WEF2025FutureJobs,Cazzaniga2024Gen,gmyrek2025genai,oecd2023employment}. These measures  now inform decisions by policymakers and firms who may not be able to verify the underlying models or methods, even as existing exposure indices disagree substantially on which occupations are most exposed \citep{yalebudgetlab2026exposure}.
As LLMs are deployed as zero-shot oracles to generate these high-stakes economic datasets, the machine learning community has a responsibility to scrutinize and correct the measurement paradigms built upon our models.

This measurement problem is urgent because the technological frontier is moving quickly. Since the release of ChatGPT in 2022, generative AI has expanded from a conversational interface into a tool embedded in research, software engineering, communication, and organizational workflows. Recent agentic systems have also widened the set of economically meaningful tasks AI could accelerate \citep{xu2025llm}. At the same time, workplace adoption remains uneven across occupations and industries, indicating that technical capability does not map mechanically into realized labor-market impact \citep{gallup2026workplaceai}. Because the frontier moves rapidly, relying on a model's training-data priors to assess real-world utility is a flawed approach to measurement.

This is a debate for the NeurIPS community because LLMs are not only used as deployed systems, but also as instruments for producing economic and policy measurements about those systems \cite{bommasani2025neurips}. When model-generated labels are reused as datasets, covariates, or policy inputs, their epistemic status changes: they become claims about latent constructs in the world that require evidence, auditing, and renewal \citep{jacobs2021measurement,blodgett2021stereotyping}. The community therefore needs standards for when ML-mediated measurements are usable and policy-ready.

\textbf{Occupational AI exposure studies should not treat zero-shot LLM judgments as valid measurement instruments unless they are externally grounded, inspectable, and periodically revalidated.} Current theoretical exposure measures fall short of this standard. \citet{eloundou2024gpts} formalized a task-level approach to measuring AI exposure and introduced the E0--E3 taxonomy that we retain for comparability. But their framework, and later extensions of it \citep{gmyrek2025genai}, classify AI task exposure with zero-shot prompts using a single proprietary model, without retrieved evidence about work content or current AI capabilities, and without a structured reasoning process. This lack of grounding introduces severe measurement fragility: in our setting, changing only the sampling temperature from 0.0 to 1.0 under a zero-shot prompt changes the assigned label for more than 20\% of occupation--task pairs. 
Our contribution is a position about evidence-based standards for ML-mediated economic measurement. We are not primarily proposing a new benchmark, dataset, or method for occupational exposure scoring. Instead, we argue that policy-facing exposure scores should not be accepted when they are produced only from ungrounded LLM priors, and we use an empirical implementation to show that the proposed standard is feasible and consequential.

To demonstrate that our proposed evidence-grounded standard is both necessary and computationally feasible, we construct a retrieval-augmented baseline for occupational exposure. Rather than treating an LLM as an ungrounded oracle, we condition open-weight reasoning models on retrieved external evidence, specifically using public news documenting real-world AI deployments and academic abstracts providing systematic capability assessments, paired with O*NET occupational descriptors \citep{onet_database_30_2}. Under this grounded protocol, disagreement across temperatures falls below 1.6\%. Relative to a zero-context baseline, the evidence-grounded condition is preferred by both LLM judges and human annotators in over 72\% of disagreement cases, and it aligns more closely with an observed usage-based exposure measure derived from real-world Claude interactions \citep{massenkoffmccrory2026labor}. 

We therefore propose that the ML community adopt three mandatory standards for occupational AI exposure measurement: external grounding,  inspectability and periodic revalidation. The pipeline in this paper is an existence proof and stress test for that position, not a claim that this particular implementation is the final exposure index. Theoretical AI exposure should not be published as a static, immutable dataset, and because AI capabilities continue to change, the measurements used to inform policy must evolve with them\footnote{Data and code are anonymized and available at \href{https://github.com/MIT-Work-Analytics-Laboratory/RAG-Exposure}{github.com/MIT-Work-Analytics-Laboratory/RAG-Exposure}}.

%% file: 2_relatedwork.tex
\label{sec:related}

\label{sec:related}

\paragraph{Task-Based AI Exposure Measurement}
Task-based approaches trace back to different works showing that tasks, rather than occupations or industries, are the relevant unit of analysis \citep{10.1257/jep.33.2.3,10.1162/003355303322552801}. Early AI exposure measures adapted this idea using patent text or cognitive-ability mappings \citep{webb2019impact,https://doi.org/10.1002/smj.3286}. Recent work has extended exposure measurement to generative AI and LLMs \citep{felten2023occupational,eloundou2024gpts}: \citet{eloundou2024gpts} classify tasks with zero-shot prompting into the E0--E3 taxonomy (defined in Appendix \ref{app:rubricdefs}), and the ILO extends the same basic approach globally \citep{gmyrek2025genai}. The IMF and OECD draw on related frameworks in macroeconomic analyses \citep{Cazzaniga2024Gen,oecd2023employment}. While these works established task-based AI exposure as a policy-relevant paradigm, they rely on zero-shot classification without external evidence or explicit reasoning.

\paragraph{Observed Usage vs. Theoretical Capability}
In contrast to theoretical capability models, recent work measures realized AI use from large-scale interaction traces. \citet{tomlinson2025working} map anonymized Bing Copilot conversations to occupational activities, while \citet{massenkoffmccrory2026labor} classify tasks observed in Claude interactions to construct an occupation-level exposure measure. We view these observed measures as crucial external validation signals. However, because observed usage is platform-specific and adoption-dependent, it is complementary to theoretical exposure measurement and not a substitute. 

\paragraph{Importing Epistemological Safeguards to Economic Measurement}
Within the broader machine learning community, retrieval-augmented generation is widely used to ground model outputs in external evidence, significantly improving factuality relative to purely parametric generation \citep{lewis2021retrievalaugmentedgenerationknowledgeintensivenlp,shuster2021retrievalaugmentationreduceshallucination,gao2024rag}. Similarly, explicit reasoning instructions have been proven to improve performance and calibration on complex judgments \citep{wei2022cot,kojima2022zeroshotcot}. Yet, these standard epistemological safeguards have been largely absent from macroeconomic AI exposure pipelines. Our position is that the techniques used to ensure factuality and inspectability in general NLP tasks must become mandatory requirements for policy-facing occupational exposure measurement.

%% file: 3_data.tex
\section{Operationalizing an Evidence-Grounded Standard}
\label{sec:data_proxies}

To transition occupational AI exposure from an ungrounded model prior to an observable measurement, a framework must explicitly define what constitutes valid external evidence. We operationalize our proposed standard using a canonical task taxonomy, temporally updated text corpora to proxy AI capabilities, and structural workplace descriptors to constrain model speculation.

\paragraph{O*NET Occupations and Tasks}
We build our framework on the O*NET 30.2 database \citep{onet_database_30_2}, which covers 923 occupations in the United States and serves as the standard taxonomy for labor economics. While this database provides a comprehensive baseline of occupation--task pairs, isolated task descriptions (e.g., Detailed Work Activities) are often semantically underspecified. When forced to evaluate these brief text strings in a zero-shot setting, LLMs may hallucinate the underlying difficulty, tools, and context of the work. To measure exposure rigorously, these bare task strings must be enriched with external context. (Sample tasks are provided in Appendix~\ref{app:exampletasks}.)

\paragraph{Evidence Proxies: News and Research Corpora}
\label{retrievaldata}
News coverage and research abstracts are noisy but informative time-stamped proxies for AI capabilities. News articles document concrete deployments, product launches, failures, and organizational uses of AI tools. Academic papers provide systematic evidence on where AI actually improves or fails to improve task performance \citep[e.g.][]{bouquet2026news,noyzhang2023productivity,brynjolfsson2025generative,dellacqua2023navigating}. Together, these corpora trace the observable, evolving boundary of what current systems can plausibly accelerate. For this, we retrieve and scrape over 34,000 news articles from 2025 and early 2026 using GDELT \citep{leetaru2013gdelt} from a predefined set of news sources (Appendix \ref{app:news_corpus}) and 19,000 scholarly abstracts \citep{Kinney2023TheSS} filtered for work at the intersection of AI, labor, and economics.

\paragraph{Anchoring Claims with Workplace Constraints}
While news and academic corpora provide crucial evidence on AI capabilities, they may suffer from tech-optimism, reporting on future potential rather than current practice. To defend against speculative exposure claims, we use occupation-level O*NET survey descriptors that summarize persistent features of the work environment. These descriptors do not serve as direct evidence of task exposure. Instead, they provide structured occupational context, such as the importance of face-to-face interaction, responsibility for others' safety, degree of automation, and precision requirements. They function as grounded priors on the work setting that force the LLM to situate each task in a realistic physical and organizational context, drastically reducing purely speculative judgments. (Appendix~\ref{app:onet_descriptors} reports the full set of survey questions).

%% file: 4_method.tex

\section{An Existence Proof for Evidence-Grounded Measurement}
\label{sec:inference_pipeline}

An open-source pipeline satisfying the proposed standards can be constructed using only open-weight models, public retrieval tools, and worker-reported data. The point of this section is not to introduce a novel retrieval architecture, but to show that reproducibility, external grounding, and inspectability are practical requirements for occupational AI exposure measurement.

\paragraph{Contextualized Retrieval as External Grounding}
Because isolated task descriptions are often too underspecified to identify relevant evidence, query expansion including the O*NET job description and associated intermediate work activities helps retrieve more specific and relevant evidence \cite{li2025query}. We then retrieve supporting evidence from our compiled corpora using a standard two-stage retrieval-augmented generation (RAG) pipeline. The first stage uses a hybrid embedding model to capture both dense semantic relevance and sparse lexical matching \cite{bge-m3,chen2024m3}, while the second stage reranks candidates using a cross-encoder to maximize precision \cite{li2023making}. This ensures that the evidence set remains compact enough for the LLM's context window while being highly relevant to the specific occupation (more details in Appendix \ref{app:method}). 

\paragraph{Open-Weight Inference for Inspectability}
Given the retrieved evidence and O*NET survey descriptors, inference is performed separately for each occupation--task pair using open-weight reasoning and instruct models (Qwen3-30B-Thinking, Ministral3-14B-Reasoning, and Gemma4-31B Instruct) \citep{yang2025qwen3,liu2026ministral,gemma4_model_card}. Using open-weight models satisfies the inspectability requirement of our proposed standard, ensuring that the reasoning traces behind economic measurements can be independently audited.

To translate these model judgments into an up-to-date, viable economic measure, we build on the E0--E3 taxonomy established by \citet{eloundou2024gpts} to preserve comparability, but update the operational meaning of the classes to reflect the 2026 AI frontier. Because contemporary systems are highly agentic and multimodal, tasks that previously required complex custom integration may now be directly exposed. The models classify each task based on the retrieved evidence:
\begin{itemize}
    \item[$E0$] \textbf{No meaningful exposure:} Direct access to a 2026 agentic AI system cannot reduce task completion time by at least 50\% at equivalent quality.
    \item[$E1$] \textbf{Direct exposure:} A 2026 agentic AI system, using only standard browser and workspace capabilities, can reduce task completion time by at least 50\%.
    \item[$E2$] \textbf{Integration-dependent exposure:} A standalone agent is insufficient, but additional software, enterprise integrations, or organization-specific tooling built on top of the agent could achieve the time reduction.
    \item[$E3$] \textbf{Vision-dependent exposure:} The task reaches the 50\% threshold only when image or video capabilities are added and visual understanding is the binding constraint.
\end{itemize}
Following prior work, these labels are converted into a discrete numerical score $\beta_{k,i} \in \{0, 0.5, 1\}$ as a proxy for the degree of task exposure (Full rubric definitions and model instructions are provided in Appendix \ref{app:rubricdefs} and more details on task exposure measurements in Appendix \ref{app:e}).

\paragraph{Occupation-Level Exposure Measurement}
The final step in exposure measurement is aggregating task-level scores into occupation-level indices. Prior approaches have often fallen back on ungrounded assumptions at this stage: \citet{eloundou2024gpts} assigned uniform weights within coarse categorical designations (effectively treating all "core" tasks as equally important), while \citet{massenkoffmccrory2026labor} relied on task-time weights elicited directly from a language model. 

Coarse weighting implicitly treats peripheral and central tasks as equally time-consuming, while model-elicited weights introduce another layer of opaque model judgment. To maintain our standard of external grounding, we follow \citet{measuringintensive} and aggregate scores using worker-reported task shares derived from O*NET survey responses. Let $\pi_{k,i}$ denote the normalized, reported task share for task $i$ in occupation $k$, with $\sum_{i} \pi_{k,i} = 1$. We define occupation-level exposure as:
\begin{equation}
\varepsilon_k = \sum_{i} \beta_{k,i}\pi_{k,i}
    \label{eq:jobexposure}
\end{equation}
By anchoring task shares in worker-reported data on how work is actually organized, this weighting scheme guarantees that tasks occupying a larger share of real-world labor exert greater influence on the final measurement.

%% file: 6_results.tex
\section{Evidence for the Position: Grounding Changes Measurement Claims}
\label{sec:results}

We evaluate our framework not to establish a new state-of-the-art benchmark, but as a testimony to our position: does access to AI capability evidence and structured reasoning fundamentally change AI exposure measurements in a direction that is more defensible and aligned with observed AI usage? (Reasoning instructions are detailed in Appendix \ref{app:reasoninginstruct})

To answer this, we evaluate the pipeline along two dimensions tied directly to the proposed standard: whether the model actually uses the evidence, and whether evidence changes the final measurement in a defensible direction. For label quality, we compare our grounded condition against a zero-context baseline. We evaluate the resulting disagreements using automatic pairwise judgments (via Prometheus 7B \citep{kim2024prometheus}, a model specialized in judging LLM outputs, and GPT-5.4 Mini \cite{openai2026gpt54mini}) and human preference annotation, following a growing literature on LLM-based evaluation and pairwise judging \citep{liu2023geval,zheng2023judging}.

\subsection{Does the Measurement Actually Use Evidence?}
Before assessing label changes, we must ensure the models are genuinely relying on the retrieved evidence rather than ignoring it in favor of their priors. We evaluate this using Prometheus 7B \citep{kim2024prometheus} to measure two properties: groundedness (whether reasoning makes conservative use of retrieved context) and faithfulness (whether the final rationale is fully supported by the provided context bundle, penalizing hallucinated facts). 

As shown in Table \ref{tab:retrieval_eval_prometheus}, scores are consistently strong across all open-weight models. This confirms that our proposed standard is operationalizable: models can successfully override their internal priors when forced to condition judgments on external evidence.

\begin{table}[!h]
\centering
\small
\caption{Prometheus 7B evaluation scores across 18{,}796 occupation--task pairs. Entries report mean \(\pm\) standard deviation on a 1--5 Likert scale \cite{likert1932technique}. The \% 4--5 columns report the share of samples receiving high ratings. Higher values indicate more faithful and better grounded reasoning.}
\begin{tabular}{lcccc}
\toprule
Model & Faithfulness & \% 4--5 & Groundedness & \% 4--5 \\
& (mean \(\pm\) sd) & & (mean \(\pm\) sd) & \\
\midrule
Qwen3 30B Thinking     & 4.29 \(\pm\) 0.55 & 97\% & 4.57 \(\pm\) 0.52 & 99\% \\
Ministral 14B Reasoning& 4.09 \(\pm\) 0.46 & 93\% & 4.32 \(\pm\) 0.56 & 95\% \\
Gemma 4 31B Instruct   & 3.91 \(\pm\) 0.71 & 85\% & 4.16 \(\pm\) 0.65 & 91\% \\
\bottomrule
\end{tabular}
\label{tab:retrieval_eval_prometheus}
\end{table}

\subsection{Does Grounding Change Labels in a Defensible Direction?}

\begin{figure}[!h]
\vspace{-12pt}
    \centering
    \maybeincludegraphics[width=\linewidth]{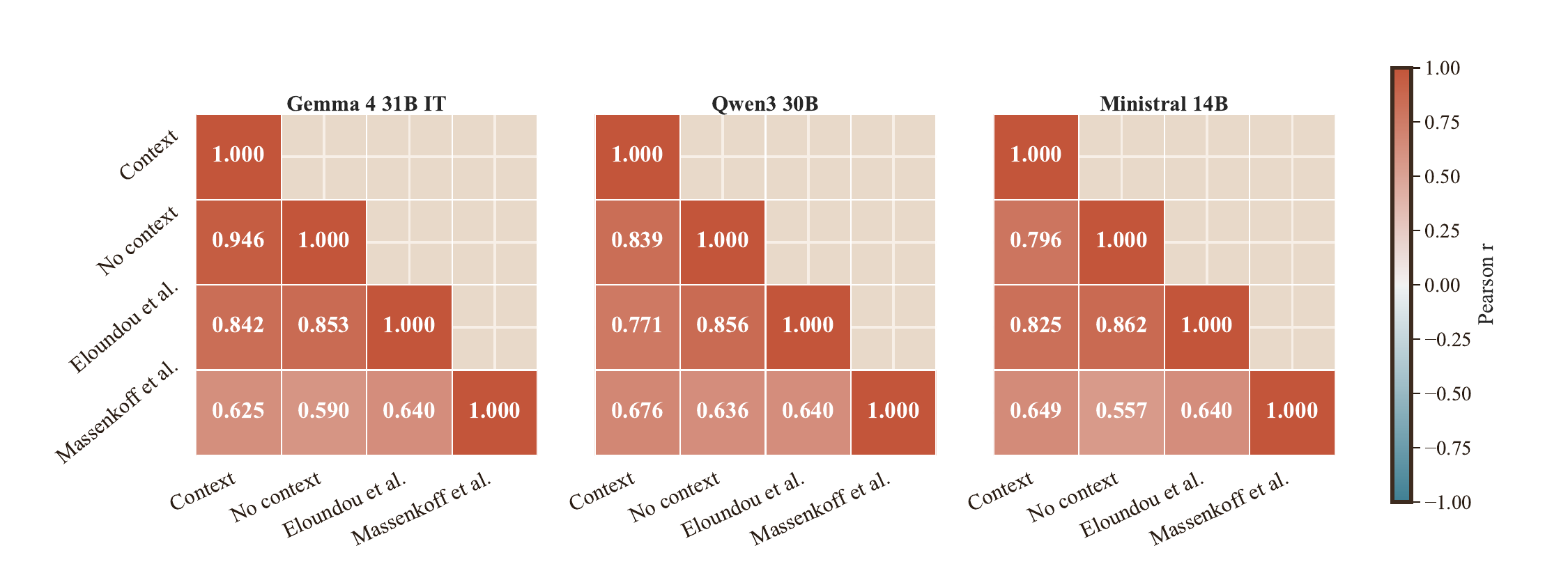}
    \caption{Pairwise Pearson correlations between occupation-level exposure scores under the context and no-context conditions, together with the occupation-level measures from \citet{eloundou2024gpts} and \citet{massenkoffmccrory2026labor}.}
    \label{fig:correlations}
\end{figure}

\paragraph{Alignment with Observed Reality.}
Figure \ref{fig:correlations} compares occupation-level exposure scores from our two inference settings against prior measures from \citet{eloundou2024gpts} (theoretical scores from 2023) and \citet{massenkoffmccrory2026labor} (observed exposure derived from real-world Claude usage). Across all three model families, the evidence-grounded condition correlates more strongly with the observed usage-based measure than the zero-context baseline does. For Gemma, the correlation increases by 3.5\%; for Qwen, by 4\%; and for Ministral, by 9.2\%.

Crucially, the context condition does not simply maximize agreement with \citet{eloundou2024gpts}. In fact, correlation with \citet{eloundou2024gpts} is slightly lower under the context setting than under the no-context baseline for all three models. We do not view this divergence as a weakness. If retrieving external evidence successfully overrides ungrounded model priors, the resulting exposure scores should not replicate prior zero-shot labels exactly. Because the \citet{massenkoffmccrory2026labor} measure reflects realized AI adoption patterns rather than ex ante capability judgments, stronger alignment with that measure provides directional evidence that our evidence-grounded standard produces more realistic estimates.

\begin{figure}[!h]
\vspace{-12pt}
    \centering
    \maybeincludegraphics[width=\linewidth]{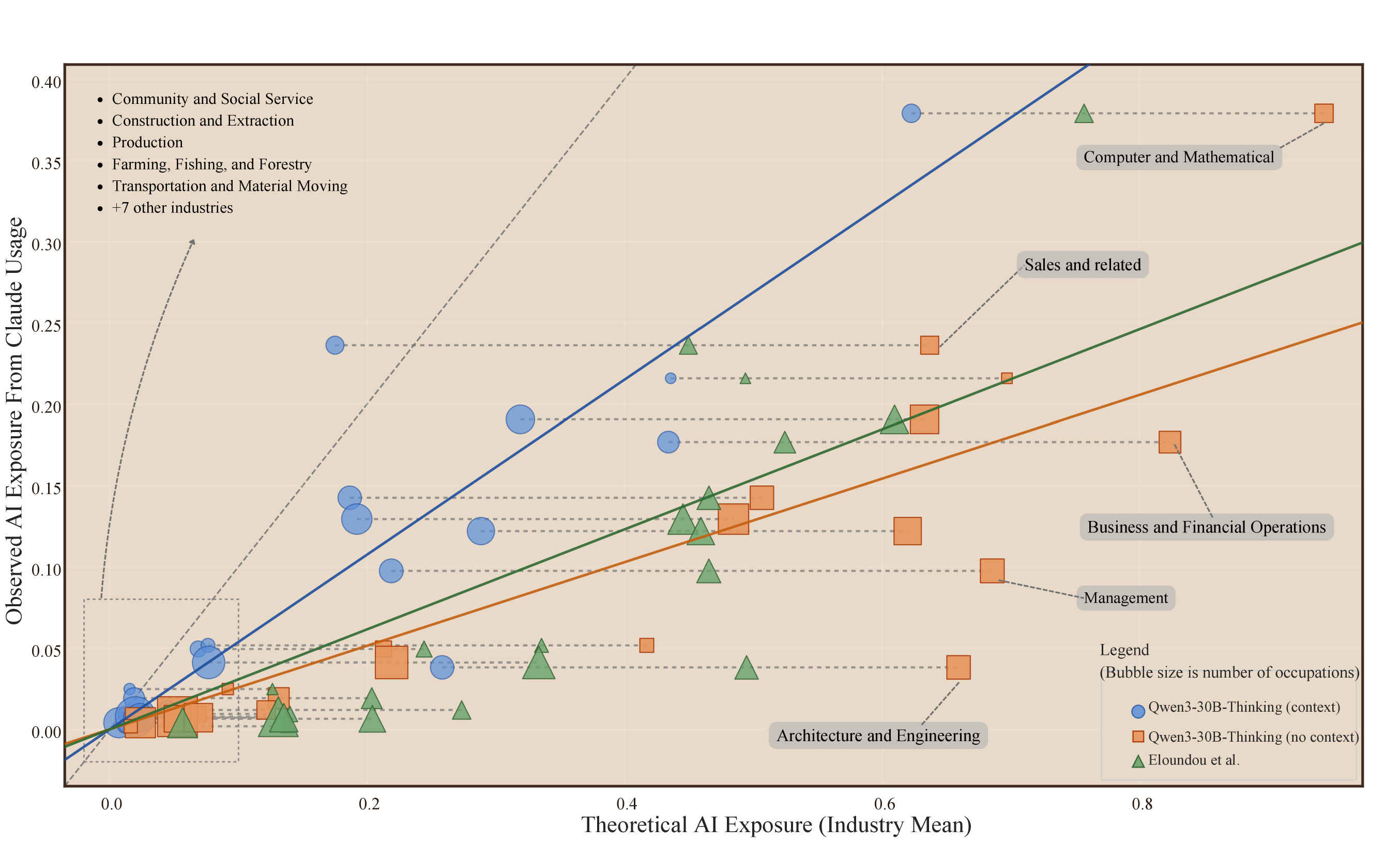}
    \caption{Industry-level alignment between theoretical AI exposure (x-axis) and observed Claude usage (y-axis). Blue bubbles represent our evidence-grounded estimates (Qwen3), orange bubbles represent an ungrounded zero-shot ablation (Qwen3 no context), and green bubbles represent prior zero-shot estimates from \citet{eloundou2024gpts}. Dotted lines connect the same industry across the three methodologies. Bubbles for each individual industry are aligned along the same y value.}
    \label{fig:industry_scatter}
\end{figure}

\paragraph{Visualizing the Grounding Effect at the Industry Level.}
To understand how evidence-grounding changes the macroeconomic picture, Figure \ref{fig:industry_scatter} plots industry-level mean theoretical exposure (x-axis) against observed real-world Claude usage (y-axis) from \citet{massenkoffmccrory2026labor}. We plot three measurements: prior zero-shot estimates from \citet{eloundou2024gpts} (green), our zero-context baseline using the updated 2026 rubric (orange), and our evidence-grounded estimates (blue).

The visualization reveals a systematic over-prediction in ungrounded methods. Both the prior estimates (green) and our updated zero-context baseline (orange) are frequently pushed far to the right of the 45-degree line, indicating theoretical scores that vastly exceed realized usage. Crucially, simply updating the rubric and using a modern model without retrieved context (orange) fails to correct this inflation; in some industries, it exacerbates it. 

By contrast, conditioning the model on retrieved evidence and O*NET constraints (blue) consistently pulls the theoretical estimates leftward, closer to observed reality. This shift is pervasive across both highly exposed sectors (e.g., ``Computer and Mathematical'', ``Business and Financial Operations'') and lower-exposure sectors (e.g., ``Management''). This pattern reinforces our core position: \textbf{without external evidence to anchor them, models acting as zero-shot oracles tend to hallucinate overly optimistic time-savings, resulting in inflated economic exposure indices. Updating the LLM or the prompt is insufficient; the measurement paradigm itself must change to require external grounding.}

\begin{figure}[!h]
\vspace{-12pt}
\centering
\maybeincludegraphics[width=\linewidth]{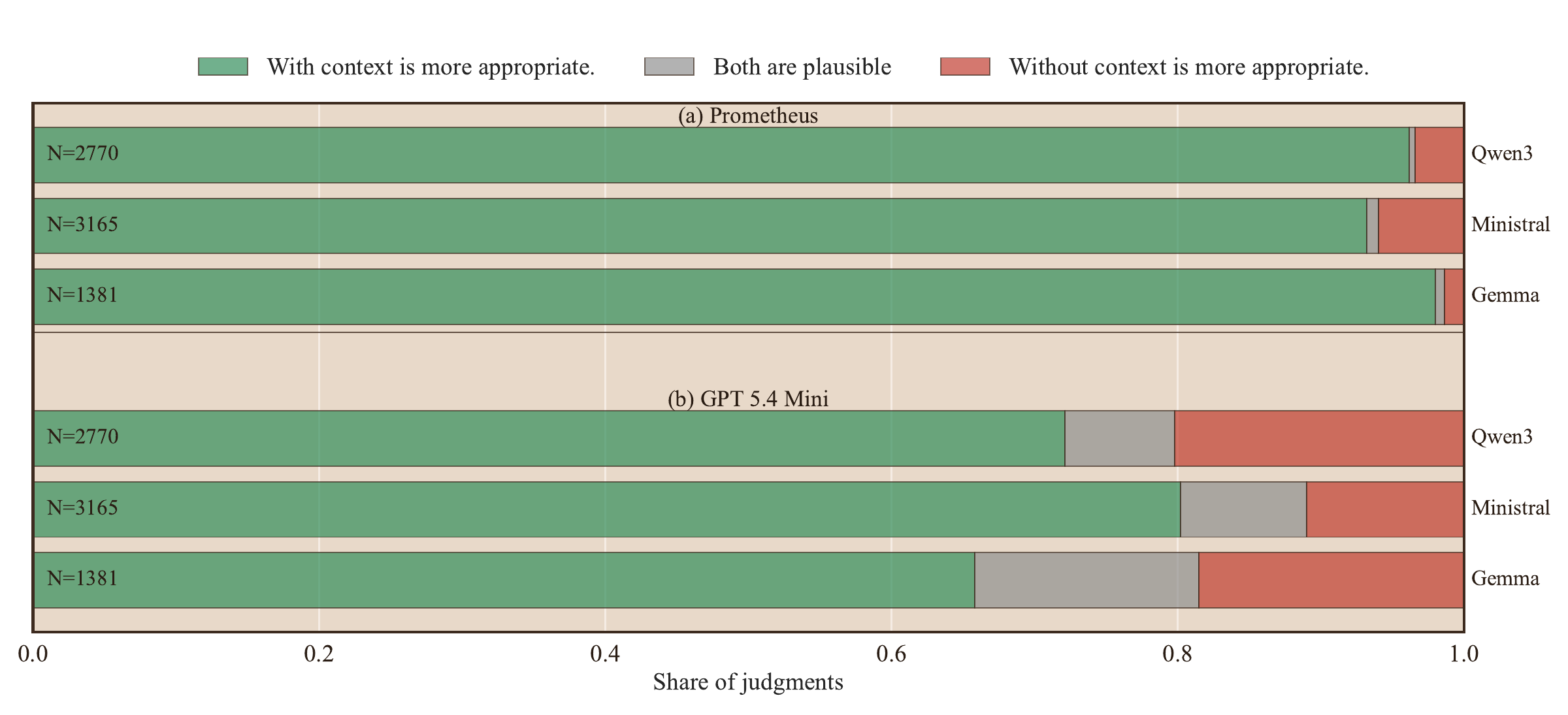}
\caption{Pairwise preference judgments on occupation–task pairs for which the context and no context conditions disagree. Bars report the share of cases in which the context grounded output is judged more appropriate, the no context output is judged more appropriate, or the two are plausible. Panel (a) uses Prometheus with access to both labels and rationales and panel (b) uses GPT 5.4 Mini with labels and rationales.}
\label{fig:1v1}
\end{figure}

\paragraph{Context vs.\ No Context Baseline Evaluations.}
To test whether access to evidence changes judgments in a preferred direction, we isolate the subset of occupation–task pairs where the grounded and zero-context settings produce different labels. Figure~\ref{fig:1v1} shows a consistent preference for the grounded condition across all three model families and evaluators. When Prometheus \cite{kim2024prometheus} is given both labels and rationales, the context-grounded output is preferred in the large majority of disagreement cases. GPT 5.4 Mini is more conservative but follows the same trend, favoring the context-grounded condition while assigning a larger share to the \textit{Both are plausible} category. This indicates that when retrieval changes the predicted label, it does so in a direction that is systematically judged to be more appropriate and better justified under the rubric. Most disagreements between the two conditions occur around the $E0$/$E1$ and $E1$/$E2$ label pairs; these transition-level disagreements are examined in Appendix \ref{app:transition_patterns}.

\paragraph{Human Evaluations}
We supplement the automatic pairwise judgments with human evaluation on 200 sampled disagreement cases spanning five diverse job families. Five annotators (authors and research group members) evaluated each sample. Annotators were shown only the two candidate exposure labels in randomized order, without rationales, to prevent stylistic anchoring. The human judgments strongly mirror the LLM evaluators: the majority vote among the five annotators selected the context-grounded label in 70.1\% of cases, compared with 24.3\% for the no-context label (with 5.6\% marked as neither). Inter-annotator agreement is moderate, with a Fleiss' $\kappa$ of 0.36, which is expected given the subjective nature of forecasting task-level AI exposure with no rationales behind it. The user interface is presented in Appendix \ref{app:human_eval_interface}.

\subsection{Exposure Distributions and the Moving Frontier}
\begin{figure}[!h]
    \centering
    \maybeincludegraphics[width=0.7\linewidth]{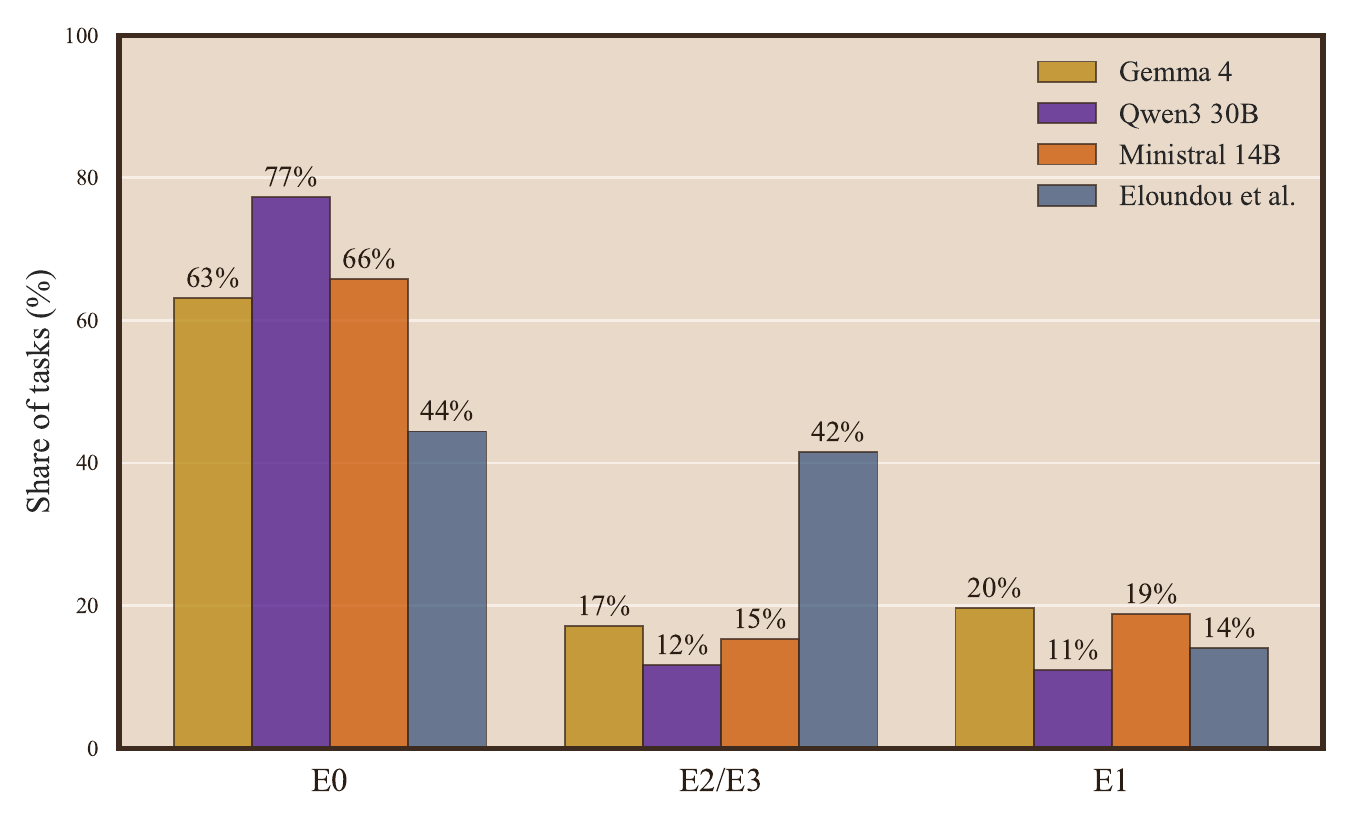}
    \caption{Task-level exposure label distributions: with-context models vs. \citet{eloundou2024gpts}.}    
    \label{fig:exposure_dist}
\end{figure}

Figure \ref{fig:exposure_dist} compares the distribution of task-level exposure labels under our grounded 2026 framework with the zero-shot GPT-4 labels reported by \citet{eloundou2024gpts}. Two shifts are apparent. First, the share of E2/E3 labels is substantially lower, reflecting a frontier effect: contemporary agentic systems can now natively browse and interact with digital tools, rendering some previously integration-dependent tasks directly exposed (E1).
Second, our models assign a substantially larger share of tasks to E0 (No Exposure). This does not imply present-day systems are weaker. Rather, it demonstrates the conservative effect of evidence grounding. Because our framework requires models to condition judgments on retrieved evidence and O*NET work-context descriptors (e.g., physical constraints, in-person requirements), it is far less likely to hallucinate time-savings in borderline cases. 

\textbf{We emphasize that historical exposure scores should be interpreted as dated measurements of a moving capability frontier, not as immutable labels. This is particularly important for policymakers and researchers, who may reuse existing exposure datasets because they are convenient, standardized, and already widely cited. Doing so without accounting for changes in model capabilities risks hard-coding outdated assumptions into present-day labor-market analysis. In our view, theoretical AI exposure should be periodically re-measured with updated rubrics and explicit evidence, rather than inherited as a static ground-truth variable.}

%% file: 7_limitations.tex
\paragraph{Discussion.} 
The central claim of this paper is a position about the standards that should govern policy-facing ML measurements: occupational AI exposure scores should be treated as observable measurement claims, not as outputs of ungrounded LLM priors. Consequently, pipelines generating these scores should satisfy three standards: external grounding, inspectability and periodic revalidation. The retrieval-augmented implementation in this paper is evidence for the feasibility and importance of that standard, not the final word on how occupational AI exposure must be scored.

A practical consequence of our proposed standard is that measurement can be kept current without waiting for a new model to finish pre-training. Because evidence sources (news and academic papers) refresh on different schedules, the inference pipeline can be rerun dynamically as capabilities and deployment contexts evolve. Theoretical exposure scores are not timeless facts about the labor market; they are frontier-dependent measurements whose validity decays rapidly, and they must be treated accordingly by policymakers. 

Our results also clarify how observed-usage measures should be interpreted. We regard the occupation-level analysis of \citet{massenkoffmccrory2026labor} as a crucial contribution that moves the literature toward realized AI use, and we use it as an external validation signal throughout this work. However, because observed usage reflects a specific platform's user population and deployment context rather than economy-wide AI adoption, it cannot replace theoretical capability measurement. Observed usage and theoretical exposure answer different questions, and are best understood as complements.

\paragraph{Limitations.} 
Our framework does not produce ground-truth exposure labels, and we do not believe such ground truth currently exists. Exposure is a latent construct that depends on task definitions, time-savings thresholds, model capabilities, and organizational context. Our claim is therefore comparative: grounding measurement in external evidence yields a more inspectable and defensible proxy than zero-shot judgment alone, but it is not a definitive answer. 

Our evidence base also remains incomplete. News coverage is uneven across occupations, and many physically embodied or less-discussed sectors have limited retrieved context at inference time (Appendix \ref{app:chunking_coverage}). Consequently, the pipeline reduces, but does not eliminate, dependence on model judgment. Furthermore, the inherited $E0$--$E3$ taxonomy is coarse, forcing borderline cases into discrete bins. Future work should explore continuous measures that separately capture dimensions such as direct task acceleration, integration dependence, and quality-risk constraints.

\paragraph{Alternative Views and Objections.} 
One objection is that this paper belongs in the main NeurIPS track because it contains a measurement pipeline and empirical evaluation. The pipeline is deliberately built from conventional components: retrieval, reranking, open-weight inference, structured rubrics, and worker-reported task weights. Its role is not to claim algorithmic novelty or establish a new benchmark as the paper's primary contribution. Its role is to make a normative position testable: if policy-facing exposure measures depend on LLM judgments, those judgments should be externally grounded, inspectable, reproducible, and periodically revalidated.

One alternate view holds that the strongest proprietary model (e.g., GPT-based models) should serve as the measurement instrument, since exposure scoring is a capability-sensitive task. While we recognize the appeal of maximizing frontier intelligence, we maintain that a measurement pipeline intended for public policy cannot depend entirely on a black-box artifact whose reasoning cannot be independently audited and whose training data remains undisclosed. Methodological inspectability must take precedence over API convenience.

Another view argues that theoretical exposure should give way entirely to observed-usage measures. While observed usage is indispensable, it is adoption-dependent and can severely understate actual AI capabilities in sectors where organizational adoption is simply slow. Realized use measures what workers \textit{are} doing; theoretical exposure measures what they \textit{could} be doing. Both are required for effective workforce planning.

\section{Conclusion}
As AI systems become increasingly agentic, governments and institutions are urgently seeking data to anticipate labor market disruptions. However, the machine learning community must not allow zero-shot LLM prompts to become the default instrument for macroeconomic measurement. This paper's contribution is not a new definitive exposure score, but a position: policy-facing AI exposure measurements should be treated as auditable measurement claims, and should not be accepted when produced only by ungrounded LLM priors.

We have demonstrated that transitioning to an evidence-grounded paradigm is both necessary and computationally feasible. By forcing open-weight reasoning models to condition their judgments on retrieved news, academic literature, and structural workplace descriptors, we produce exposure scores that are highly stable, auditable, and better aligned with observed real-world adoption. To build economic data we can trust, we must treat AI exposure not as a zero-shot prediction, but as a claim that demands evidence and periodic remeasurement.

%% file: 8_appendix.tex
\appendix

\section{Rubric and Model Inputs}
\label{app:labeling_framework}

This appendix provides supplementary material for the labeling framework, retrieval corpus construction, and evaluation procedures used in the main paper. We first document the exposure rubric and model inputs, then report additional retrieval details, evaluation materials, and supplementary observational figures.

\subsection{Rubric Definitions}
\label{app:rubricdefs}

We retain the four-class exposure structure of Eloundou et al.~\citep{eloundou2024gpts} for comparability, but update the operational meaning of each class to reflect the 2026 agentic AI frontier described in Appendix \ref{app:rubricdefs}. Table~\ref{tab:exposure_taxonomy_2026} reports the full rubric definitions used during inference.

\begin{table}[htbp]
\centering
\caption{Exposure taxonomy for occupational tasks in the presence of a 2026 agentic AI system.}
\label{tab:exposure_taxonomy_2026}
\begin{tabularx}{\textwidth}{p{0.05\textwidth} X}
\hline
\textbf{Class} & \textbf{Definition} \\
\hline

$E0$ &
\textbf{No exposure.} Label tasks as $E0$ if direct access to the 2026 agentic chatbot interface cannot reduce the time required to complete the task by at least 50\% at equivalent quality. This class also includes tasks where more than half of the work is inherently physical, embodied, or dependent on in-person human interaction that cannot be substituted by text, workspace actions, or browser-based actions. It further includes tasks where the binding constraint is access to a physical environment, regulated sign-off that must occur in person, or real-time interpersonal dynamics such as speeches or instruction-giving that cannot be meaningfully accelerated through drafting, planning, or digital coordination alone. \\[0.6em]

$E1$ &
\textbf{Direct exposure.} Label tasks as $E1$ if a 2026 agentic chatbot, without any specialized enterprise integrations beyond standard browser and workspace access, can reduce completion time by at least 50\% at equivalent quality. This assumes the agent operates through general-purpose capabilities such as text generation, reasoning, multi-step planning, browsing public or credential-accessible websites, and interacting with standard productivity software. Typical examples include writing, coding, translation, summarization, structured web research, spreadsheet or document work, routine browser-based administration, and planning or coordination tasks where the agent's standalone capabilities are sufficient to achieve the time reduction. \\[0.6em]

$E2$ &
\textbf{Exposure via LLM-powered applications.} Label tasks as $E2$ if the standalone 2026 agentic chatbot may not by itself reduce completion time by at least 50\%, but additional software, deeper integrations, or organization-specific tooling built on top of the agent plausibly could. The key distinction from $E1$ is that the primary bottleneck lies in reliable and structured access to proprietary systems, automated interaction with internal databases, long-horizon monitoring, compliance-constrained workflows, high-assurance environments, or tight coupling with internal business logic. In such cases, the standalone agent may still assist with reasoning, drafting, or partial workflow support, but the decisive acceleration depends on integration beyond a generic browser-based agent. \\[0.6em]

$E3$ &
\textbf{Exposure given image capabilities.} Suppose the worker also has access to an integrated system capable of viewing, captioning, and generating images, including reading scanned PDFs, extracting text from images, interpreting diagrams, and analyzing video inputs. Label tasks as $E3$ only when these image capabilities, in addition to the agentic chatbot, enable a time reduction of at least 50\% and visual understanding is the binding constraint for that reduction. This label should be used only when text-only capabilities are insufficient and visual extraction, interpretation, or generation is the critical enabler. \\

\hline
\end{tabularx}
\end{table}

\subsection{Example Occupation--Task Inputs}
\label{app:exampletasks}

Because retrieval and labeling are performed at the occupation--task level, it is useful to illustrate the structure of the inputs used by the pipeline. Table~\ref{tab:selected_tasks} reports representative examples showing the occupation title, associated intermediate and detailed work activities, and the task description provided to the model.

\renewcommand{\arraystretch}{1.25}

\begin{table}[ht]
\centering
\small
\setlength{\tabcolsep}{6pt}
\rowcolors{2}{lightgrayrow}{white}
\caption{Selected occupation--task inputs, including occupation title, intermediate work activity (IWA), detailed work activity (DWA), and task description.}
\resizebox{\linewidth}{!}{%
\begin{tabular}{p{3.2cm} p{4.4cm} p{4.8cm} p{5.4cm}}
\toprule
\rowcolor{headergray}
\textbf{Occupation Title} & \textbf{IWAs} & \textbf{DWAs} & \textbf{Task Description} \\
\midrule
Data Scientists & Analyze scientific or applied data using mathematical principles. & Apply mathematical principles or statistical approaches to solve problems in scientific or applied fields. & Propose solutions in engineering, the sciences, and other fields using mathematical theories and techniques. \\

Political Scientists & Prepare legal or regulatory documents. & Prepare information or documentation related to legal or regulatory matters. & Write drafts of legislative proposals, and prepare speeches, correspondence, and policy papers for governmental use. \\

Baristas & Prepare foods or beverages. & Cut cooked or raw foods. & Slice fruits, vegetables, desserts, or meats for use in food service. \\

Psychology Teachers, Postsecondary & Evaluate patient or client condition or treatment options. & Evaluate patient functioning, capabilities, or health. & Provide clinical services to clients, such as assessing psychological problems and conducting psychotherapy. \\
\bottomrule
\end{tabular}%
}
\label{tab:selected_tasks}
\end{table}

\subsection{O*NET Descriptor Questions}
\label{app:onet_descriptors}

In addition to retrieved external evidence, the model receives occupation-level O*NET descriptors that characterize persistent features of the work environment. Table~\ref{tab:onet_descriptors} lists the descriptor questions included in the prompt together with the associated response scales.

\begin{table}[ht]
\centering
\small
\setlength{\tabcolsep}{8pt}
\rowcolors{2}{lightgrayrow}{white}
\caption{O*NET descriptor questions and associated response scales.}
\resizebox{0.85\linewidth}{!}{%
\begin{tabular}{p{0.48\textwidth} p{0.44\textwidth}}
\toprule
\rowcolor{headergray}
\textbf{Question} & \textbf{Range of values} \\
\midrule
How frequently does your job require face-to-face discussions with individuals and within teams? &
\(100\): Every day; \(75\): Once a week or more but not every day; \(50\): Once a month or more but not every week; \(25\): Once a year or more but not every month; \(0\): Never \\

How much responsibility is there for the health and safety of others in this job? &
\(100\): Very high responsibility; \(75\): High responsibility; \(50\): Moderate responsibility; \(25\): Limited responsibility; \(0\): No responsibility \\

How automated is the job? &
\(100\): Completely automated; \(75\): Highly automated; \(50\): Moderately automated; \(25\): Slightly automated; \(0\): Not at all automated \\

How important is being very exact or highly accurate in performing this job? &
\(100\): Extremely important; \(75\): Very important; \(50\): Important; \(25\): Fairly important; \(0\): Not important at all \\

How important are continuous, repetitive, physical activities (like key entry) or mental activities (like checking entries in a ledger) to performing this job? &
\(100\): Extremely important; \(75\): Very important; \(50\): Important; \(25\): Fairly important; \(0\): Not important at all \\

How frequently does your job require written letters and memos? &
\(100\): Every day; \(75\): Once a week or more but not every day; \(50\): Once a month or more but not every week; \(25\): Once a year or more but not every month; \(0\): Never \\
\bottomrule
\end{tabular}%
}
\label{tab:onet_descriptors}
\end{table}

\subsection{Reasoning Instructions}
\label{app:reasoninginstruct}

In addition to the retrieved evidence, we provide the model with explicit reasoning instructions designed to make task-level judgments more structured and inspectable. For each occupation--task pair, the model is asked to decompose the task into a small number of concrete subtasks, assign exposure levels to those subtasks, and then synthesize them into a single overall label for the task.
The prompt explicitly instructs the model to ground its reasoning in the retrieved evidence whenever relevant, while treating the O*NET survey descriptors as occupation-level background context rather than as direct evidence of task exposure. When no relevant external evidence is retrieved, the model is instructed to state that retrieved context is not informative for the task and to rely on its task decomposition and rubric-based reasoning alone. Responses are constrained to a structured JSON format containing a short rationale, an indicator of whether retrieved context was relevant, the supporting source identifiers, and the final exposure label.

\section{Supplementary Evaluation Materials}
\label{app:supp_eval}

This section provides additional material for the evaluation procedures described in Section~5.2. We include the human annotation interface and a transition-level view of disagreement patterns between the grounded and no-context conditions.

\subsection{Transition-Level Disagreement Patterns}
\label{app:transition_patterns}
\begin{figure}[H]
    \centering
    \includegraphics[width=0.75\linewidth]{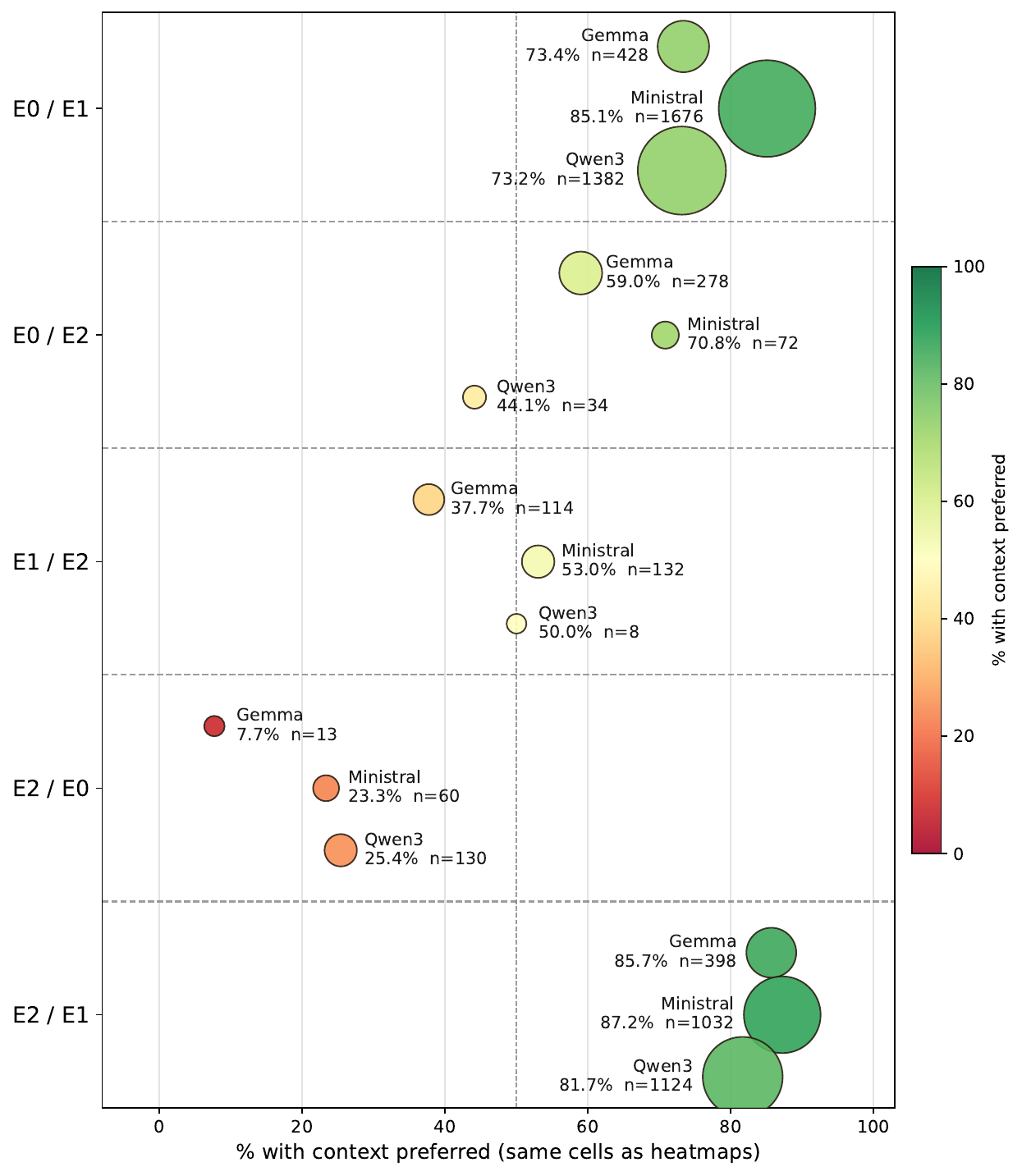}
    \caption{Transition-specific pairwise preferences on disagreement cases. Each bubble corresponds to a label transition between the no-context and context conditions. For transitions $x /y$ this denotes the context condition predicted $x$ and the no-context $y$. The color gradient shows the proportion of disagreements for each transition, and when the with context condition is preferred.}
    \label{fig:bubbles}
\end{figure}

To better understand which label changes drive the aggregate preference for the grounded condition, we decompose disagreement cases by transition type between the no-context and context settings. Figure~\ref{fig:bubbles} shows that disagreements are concentrated in a small number of transitions. Across models, the most frequent disagreements are those in which the no-context condition assigns $E1$ while the grounded condition assigns either $E0$ or $E2$. These two transition families account for the large majority of disagreement cases. In both families, the judge most often prefers the grounded label, with preference rates typically above 70\% and reaching the mid-80\% range for several model-transition pairs. This pattern suggests that retrieval does not induce arbitrary relabeling. Rather, it changes judgments primarily in a limited set of borderline cases, and when it does so, the resulting label is usually judged more appropriate under the rubric.
This pattern is consistent with the interpretation that, in the absence of retrieved context, models tend to default more often to direct-exposure judgments in borderline cases, whereas grounding with external evidence and occupational descriptors more often shifts those cases toward labels that judges consider better justified.

\subsection{Which jobs are most exposed?}
\label{app:most_exposed}
Table \ref{tab:ref_occ} reports reference occupations with high observed Claude usage together with the corresponding occupation-level theoretical exposure scores from our grounded framework and from Eloundou et al.~\citep{eloundou2024gpts}. Several occupations remain consistently highly exposed across both observed usage and our grounded theoretical estimates, including computer programmers, market research analysts and marketing specialists, financial and investment analysts, and software quality assurance analysts and testers. These occupations are dominated by text-rich, analytical, and software-mediated tasks for which current frontier systems can plausibly deliver substantial time savings.

At the same time, the grounded estimates are often more conservative than the earlier zero-shot benchmark. This pattern is especially visible for occupations such as data entry keyers, information security analysts, and software quality assurance analysts and testers, where our scores are systematically lower than the Eloundou et al. values. We interpret this gap as evidence that grounding reduces unsupported direct-exposure judgments in cases where performance depends on workflow integration, high-assurance verification, or organizational context not captured by task text alone.

Importantly, the grounded framework does not simply lower exposure uniformly. For some occupations, such as market research analysts and marketing specialists, and in some model variants customer service representatives, the grounded scores are comparable to or higher than the \citet{eloundou2024gpts} benchmark. This suggests that retrieval changes the composition of exposure judgments rather than merely shrinking them, preserving high exposure where current evidence supports it while moderating claims that are less well grounded.
\begin{table}[!h]
  \centering
  \small
  \setlength{\tabcolsep}{4.2pt}
  \renewcommand{\arraystretch}{1.08}
  \begingroup
  \definecolor{lightgrayrow}{gray}{0.92}
  \rowcolors{3}{lightgrayrow}{white}
  \caption{Occupations with the most observed AI exposure according to Claude usage \citep{massenkoffmccrory2026labor}. \textbf{Theoretical AI exposure} columns report scores from \cite{eloundou2024gpts} and \textbf{$\pi$-weighted} occupation-level scores (Equation \ref{eq:jobexposure}) for Gemma, Qwen and Ministral models. }
  \label{tab:ref_occ}
  {%
  \begin{tabular}{@{}>{\raggedright\arraybackslash}p{5.3cm}
                  >{\centering\arraybackslash}p{1.5cm}
                  >{\centering\arraybackslash}p{1cm}
                  >{\centering\arraybackslash}p{0.92cm}
                  >{\centering\arraybackslash}p{1.08cm}
                  >{\centering\arraybackslash}p{2cm}@{}}
    \toprule
    \textbf{Occupation} & \textbf{Observed} & \multicolumn{4}{c}{\textbf{Theoretical AI exposure}} \\
    & \textbf{(\%)} & \textbf{Gemma} & \textbf{Qwen} & \textbf{Ministral} & \textbf{Eloundou et al.} \\
    \cmidrule(lr){2-2}\cmidrule(lr){3-6}
    Computer Programmers & 74.5\% & 88.3\% & 78.1\% & 88.7\% & 95.0\% \\
    Customer Service Representatives & 70.1\% & 50.0\% & 53.9\% & 87.9\% & 56.8\% \\
    Data Entry Keyers & 67.1\% & 51.5\% & 39.4\% & 48.5\% & 89.3\% \\
    Medical Records Specialists & 66.7\% & 47.1\% & 58.8\% & 58.8\% & 61.8\% \\
    Market Research Analysts and Marketing Specialists & 64.8\% & 78.6\% & 75.4\% & 83.5\% & 52.5\% \\
    Medical Transcriptionists & 63.6\% & 71.1\% & 53.9\% & 66.0\% & 87.5\% \\
    Sales Representatives, Wholesale and Manufacturing, Except Technical and Scientific Products & 62.8\% & 29.8\% & 29.6\% & 36.7\% & 62.1\% \\
    Database Architects & 57.9\% & 76.2\% & 65.1\% & 61.5\% & 91.1\% \\
    Financial and Investment Analysts & 57.2\% & 67.3\% & 71.2\% & 76.9\% & 46.2\% \\
    Software Quality Assurance Analysts and Testers & 51.9\% & 73.7\% & 65.8\% & 79.2\% & 87.7\% \\
    \bottomrule
  \end{tabular}%
  }
  \endgroup
\end{table}


\subsection{Human Evaluation Interface}
\label{app:human_eval_interface}

\begin{figure}[h]
    \centering
    \includegraphics[width=0.8\linewidth]{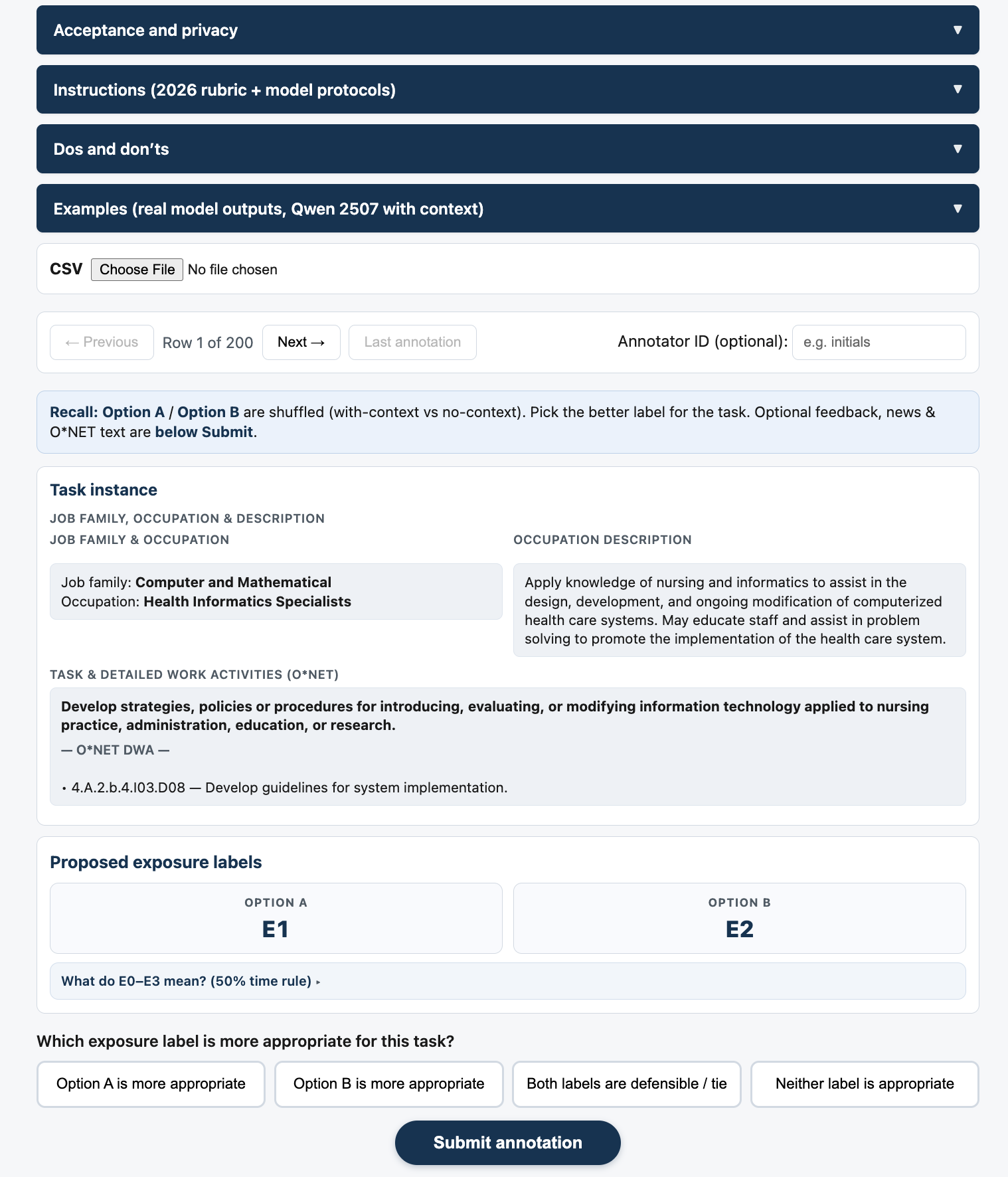}
    \caption{Human annotation interface used for pairwise evaluation on disagreement cases. Annotators were shown the occupation context, task description, and two candidate exposure labels in randomized order, without rationales.}
    \label{fig:humanevalhtml}
\end{figure}

Figure~\ref{fig:humanevalhtml} shows the annotation interface used for human pairwise evaluation. Annotators were shown the task description, occupation context, and two candidate labels in randomized order, without rationales.

\section{News Corpus}
\label{app:retrieval_details}

\subsection{News Source Distribution}
\label{app:news_corpus}

Table~\ref{tab:news-raw-scraped-counts} summarizes the composition of the news corpus by source. For each outlet, we report both the number of URLs collected from GDELT and the number of articles that were successfully scraped for downstream retrieval. \\[5pt]
\begin{table}[ht]
\makeatletter
\@ifundefined{newsrawscrapedmidsetup}{%
  \newif\ifnewsrawscrapedmidvrule
  \def\newsrawscrapedmidsep{%
    \ifnewsrawscrapedmidvrule
      \kern0.45em
      \vrule \@width 0.5\p@
        \@height \arraystretch\ht\strutbox
        \@depth \arraystretch\dp\strutbox
      \kern0.45em
    \else
      \hspace{1.2em}%
    \fi
  }%
  \global\let\newsrawscrapedmidsetup\relax
}{}%
\makeatother
\newsrawscrapedmidvruletrue
\centering
\small
\setlength{\tabcolsep}{8pt}
\rowcolors{2}{lightgrayrow}{white}
\caption{Article counts by news source, showing collected URLs from GDELT and successfully scraped articles. Scraping failures arise for several reasons, most commonly paywalls or site-specific access restrictions.}
\resizebox{0.9\linewidth}{!}{%
\begin{tabular}{lrrlrr}
\toprule
\rowcolor{headergray}
\textbf{Source} & \textbf{Raw} & \textbf{Scraped} & \textbf{Source} & \textbf{Raw} & \textbf{Scraped} \\
\midrule
yahoo.com & 14,861 & 11,854 & forbes.com & 10,958 & 8,006 \\
techradar.com & 2,222 & 1,760 & businessinsider.com & 1,345 & 1,121 \\
computerweekly.com & 1,155 & 890 & techcrunch.com & 1,037 & 875 \\
siliconangle.com & 1,042 & 860 & fortune.com & 950 & 711 \\
theregister.com & 840 & 707 & zdnet.com & 1,149 & 694 \\
cnbc.com & 831 & 672 & theguardian.com & 870 & 614 \\
theglobeandmail.com & 693 & 496 & theconversation.com & 631 & 493 \\
newsweek.com & 608 & 480 & scmp.com & 479 & 477 \\
govtech.com & 588 & 475 & theverge.com & 591 & 455 \\
nypost.com & 470 & 397 & independent.co.uk & 421 & 335 \\
foxnews.com & 422 & 295 & cnn.com & 828 & 207 \\
wired.com & 207 & 203 & arstechnica.com & 198 & 197 \\
venturebeat.com & 444 & 191 & bbc.com & 281 & 186 \\
cbsnews.com & 204 & 155 & thehill.com & 175 & 152 \\
sciencedaily.com & 170 & 151 & nature.com & 184 & 121 \\
nbcnews.com & 168 & 107 & technologyreview.com & 185 & 106 \\
time.com & 164 & 91 & bloomberg.com & 94 & 80 \\
npr.org & 175 & 80 & Others & 891 & 156 \\
\midrule
\newsrawscrapedmidvrulefalse
\textbf{Total} & \textbf{46,531} & \textbf{34,850} & & & \\
\bottomrule
\end{tabular}%
}
\label{tab:news-raw-scraped-counts}
\end{table}
%
%

%
%
\subsection{Chunking and Retrieved Context Coverage} 
\label{app:chunking_coverage}

\begin{figure}[!ht]
    \centering
    \includegraphics[width=\linewidth]{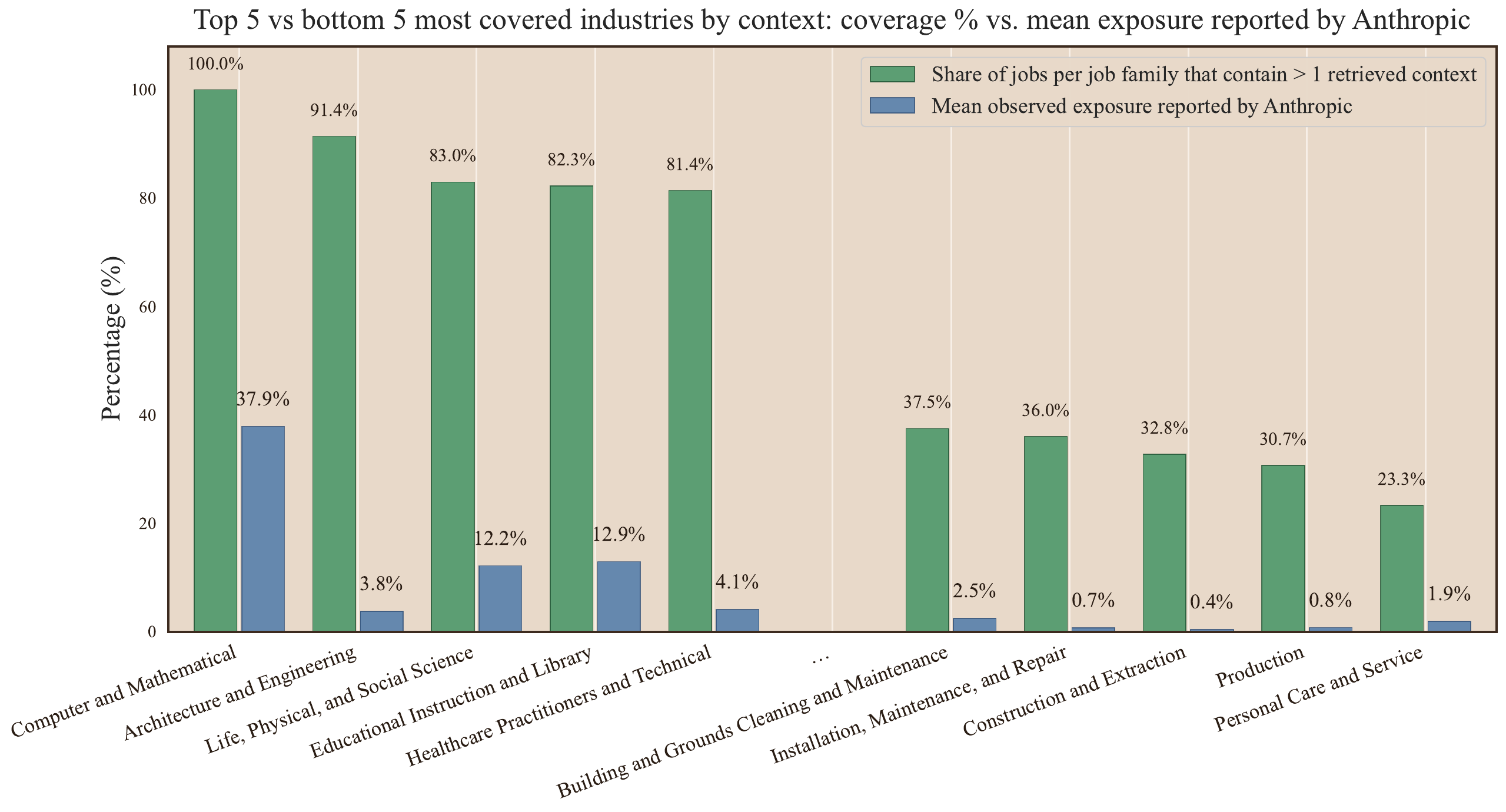}
    \caption{Comparison of the five most covered and five least covered job families by retrieved context coverage, together with mean observed exposure reported by Claude.}
    \label{fig:contextcoverage}
\end{figure}

We also report an additional diagnostic related to retrieval quality and evidence availability. 
Figure~\ref{fig:contextcoverage} reports how often retrieved context is available across job families.
As our framework relies on retrieved evidence to produce task-level exposure labels, it is important to report how often relevant context is actually available at inference time. From the expanded queries defined, we retrieve data as described in Appendix~\ref{hybridretrieval} using a minimum relevance threshold. Under this procedure, approximately 60\% of occupations have at least one task with retrieved context at inference time. Coverage is substantially higher in digitally intensive occupations and lower in sectors such as Production, Construction, and related physically embodied job families, where AI-focused reporting is comparatively sparse.

\section{Detailed Methodology}
\label{app:method}
\subsection{Retrieval Corpus and Indexing}
\label{sec:indexing}

Our retrieval corpus combines news articles and academic paper abstracts. We do not index full news articles directly, because they are often long, cover multiple topics, and contain substantial text unrelated to a given occupation--task query. Instead, we segment news articles into semantically coherent chunks and retrieve evidence at the chunk level. This allows the retriever to focus on locally relevant passages rather than entire documents. Because academic abstracts are already compact summaries, we index them without additional chunking.

After preprocessing, each indexed text unit, either a news chunk or a paper abstract, is encoded using \texttt{BAAI/bge-m3}. We use this model because it supports hybrid retrieval within a single framework, producing both dense semantic representations and sparse lexical representations. Dense representations help capture semantic similarity even when wording differs, while sparse representations preserve exact term overlap that is often useful for task- and occupation-specific retrieval. The resulting index therefore supports both semantic matching and lexical precision.

\subsection{Query Construction}
\label{sec:query_def}

Retrieval is performed at the occupation--task level. A task description on its own is often too short or underspecified to retrieve useful evidence, so we enrich each query with additional occupational context from O*NET. In particular, we construct each query by combining the occupation title, the occupation-level job description, the associated intermediate and detailed work activities, and the task text itself.

This query expansion step is important because it situates each task in its actual work setting. Rather than retrieving documents based on a short task string alone, the retriever can use the broader occupational context to identify evidence that is more specific and more relevant to the underlying work.

\subsection{Retrieval and Reranking}
\label{hybridretrieval}

We use a two-stage retrieval pipeline. In the first stage, we perform hybrid retrieval with \texttt{BAAI/bge-m3}, combining dense semantic matching with sparse lexical matching. The dense component helps identify passages that are conceptually related to the occupation--task query, while the sparse component rewards passages that contain directly relevant terms. Together, these two signals allow the retriever to balance recall and precision.

In the second stage, we rerank the top candidates using the cross-encoder \texttt{BAAI/bge-reranker-v2-m3}. Unlike the first-stage retriever, which scores queries and passages through precomputed representations, the cross-encoder jointly processes each query--passage pair and can therefore capture finer-grained relevance. We retain a small set of the highest-ranked passages as evidence for downstream inference.

\section{Moving from Task to Job Exposure}
\label{app:e}
Given the retrieval pipeline defined in Section \ref{hybridretrieval}, inference is performed separately for each occupation--task pair. For each query, we retain the top \(n=6\) retrieved passages, yielding a compact evidence set whose total length remains well within the models' context windows. These retrieved passages are supplemented with the O*NET survey descriptors from Table \ref{tab:onet_descriptors}, which provide occupation-level background information on the work environment.
We run inference with the vLLM engine \citep{kwon2023efficient}, which enables efficient large-scale serving, using open-weight reasoning-oriented and instruct models, including Qwen3 (30B Thinking) \citep{yang2025qwen3}, Ministral (14B Reasoning) \citep{liu2026ministral} and Gemma 4 (31B Instruct) \citep{gemma4_model_card}.

\paragraph{Task-Level Exposure Scores.}
\label{tasklevelscores}

For each occupation--task pair, the model assigns one of four exposure labels. Building on the taxonomy of \citet{eloundou2024gpts} with \(E0\), \(E1\), \(E2\), and \(E3\) labels, we are able to preserve comparability with prior work. We update the operational meaning of these classes to reflect the 2026 AI frontier according to definitions in Appendix \ref{app:rubricdefs}. Labels are converted into a discrete numerical task-level exposure score $\beta$:
\[
\beta = \begin{cases}
    0  & \text{if } E0 \\
    1 & \text{if } E1 \\
    0.5 & \text{if } E2 \text{ or } E3
\end{cases}
\]
\textbf{This mapping is intended as a simple proxy for the degree of task exposure}. Tasks labeled \(E1\) receive a value of 1 because they are directly exposed under current frontier systems. Tasks labeled \(E2\) or \(E3\) receive a value of 0.5 because their exposure is more conditional. These tasks may become substantially faster with additional software integration, enterprise tooling, or visual capabilities, but they are not as immediately exposed as \(E1\) tasks under current out-of-the-box systems. Tasks labeled \(E0\) receive a value of 0, indicating no meaningful exposure under the rubric.

To construct occupation-level exposure, we aggregate task-level scores within each occupation using empirically grounded task weights. We follow \citet{measuringintensive} and use task shares derived from O*NET survey responses on task relevance and frequency. Conditional on task relevance, surveyees respond to how often the task is performed. These responses are then converted into within-occupation task shares that sum to one and proxy how labor input is allocated across tasks.
O*NET-based task shares are anchored in worker-reported data on how work is actually organized, making them a more realistic and empirically grounded proxy for the intensive margin of work.
Let \(\pi_{k,i}\) denote the normalized task share for task \(i\) in occupation \(k\), with \(\sum_{i \in \mathcal{T}_k} \pi_{k,i} = 1\). We then define occupation-level exposure using Equation \ref{eq:jobexposure}. \textbf{ This weighting assigns greater influence to tasks that occupy a larger share of work in practice}.